\documentclass{article}
\usepackage[utf8]{inputenc}
\usepackage{graphicx}
\usepackage{xcolor}
\usepackage{amsmath}

\begin{document}
\title{ARA : Aggregated RAPPOR and Analysis for Centralized Differential Privacy}
\date{}
%
 
\author{Sudipta Paul \\ Junior Research Fellow, \\ School of Computer Sciences, NISER \\Bhubaneswar - 752050, Odisha, India, \\ email : sudiptapaulvixx@niser.ac.in 
\and 
\\ Subhankar Mishra \\ Assistant Professor, \\School of Computer Sciences, \\NISER Bhubaneswar - 752050, Odisha, India,\\ Homi Bhabha National Institute, \\ Training School Complex, \\ Anushaktinagar, Mumbai 400 094, India \\ email : smishra@niser.ac.in
}

\maketitle 

%
%

%
%
\begin{abstract}
    Differential privacy(DP) has now become a standard in case of sensitive statistical data analysis. The two main approaches in DP is local and central.  Both the approaches have a clear gap in terms of data storing,amount of data to be analyzed, analysis, speed etc. Local wins on the speed. We have tested the state of the art standard RAPPOR which is a local approach and supported this gap. Our work completely focuses on that part too. Here, we propose a model which initially collects RAPPOR reports from multiple clients which are then pushed to a Tf-Idf estimation model. The Tf-Idf estimation model then estimates the reports on the basis of the occurrence of "on bit" in a particular position and its contribution to that position. Thus it generates a centralized differential privacy analysis from multiple clients. Our model successfully and efficiently analyzed the major truth value every time.   
\end{abstract}
\section{Introduction}
\label{intro}
'Privacy' \cite{privacy_definition} is defined as - "a sweeping concept, encompassing (among other things) freedom of thought, control over one’s body, solitude in one’s home, control over personal information, freedom from surveillance, protection of one’s reputation, and protection from searches and interrogations". 
 Privacy is a term used to describe an individual's anonymity and how safe they feel in a location preferably in Internet, which is one of the most sensitive and concerned 'concept' at present. In the current situation crowd sourcing is the most popular source of collecting data directly from people for many research topics. Generally it is being done through several online site or portal in Internet. But there are some basic issues regarding the whole survey process, like (a) the system of survey should be convincing enough to gain the participants trust ,(b) the  processes after the survey should be effective enough to ensure the ‘truthfulness’ of the participants to the researchers, (c) The processes of research should be robust enough to guarantee the leak proof of the research architecture model, (d) The system of survey should still produce a ‘good result’ in terms of gaining an insight of the problem in spite of the ‘noise’ in the data. 
Therefore it promises a large research field in case of statistical databases where a leak of small amount of data may lead to a personal identification which might be a concern for that person in his or her personal life . The phenomenon of 'Re-Identification' was considered as a possibility at a time. However it has been a wrong solution on its own. The re-identification approach has been proved wrong by Lataniya Sweeney. She identified the medical information of the then governor of Massachusetts- William Weld, using the publicly available insurance data after cross checking it with the de-identified data published by the state govt. in the year 1997. She mailed the same data to the governor too. 
 A probable solution regarding this phenomenon was given by Lataniya Sweeney in her k-anonymity theorem\cite{Sweeney2001k-anonymity}. However, k-anonymization does not include any randomization and attackers can still make inferences about data sets that may harm individuals.  Hence, the arrival of \textit{'differential privacy'} (DP). According to Cynthia Dwork \cite{dwork2014algorithmic}, in general - " The outcome of any analysis is essentially equally likely independent of whether any individual joins or refrains from joining the dataset." Therefore it implies that if a data analyst will not know anything new about a person from a dataset of sensitive data when he/she is not present in that dataset, as well as when he/she join the dataset too in case if the dataset is maintaining the notion of DP.
\par There are two types of differential privacy - \textit{Centralized differential privacy} or CDP and \textit{Local differential privacy} or LDP. They have their own pros and cons in terms of data collection, data reservation, presence of \textit{trusted third party data curator} etc. The detail discussion and comparison between LDP and CDP is discussed in the section \ref{sec:2}.\par There are some industry standard like \textit{RAPPOR}\cite{erlingsson2014rappor} in \textit{Google}, \textit{Apple's Differential privacy}\cite{tang2017privacy} etc. in case of LDP. Also in case of CDP there is \textit{ESA architecture and its PROCHLO implementation}. \textit{RAPPOR} is fast where \textit{PROCHLO} is expensive to implement. In CDP the centralized part helps the whole approach to be compatible with existing, standard software engineering processes which is in succession helpful to use the industry standard known techniques(differentially- private releases) to analyze\cite{bittau2017prochlo} it. So, there is a clear gap between LDP and CDP approach, which is definitely a great research area.

\par There has been some approaches regarding this area like - OUTIS \cite{chowdhury2019outis}, Amplification by shuffling: from local to central differential privacy via anonymity \cite{erlingsson2019amplification} etc. OUTIS claims that it provides a bridge between LDP and CDP. OUTIS provides an architecture for differential privacy that does not need any trusted third party data curator like the CDP but still achieves the accuracy guarantees and algorithmic expressibility like in the CDP approach, which gives the possibility of "best of both the world" in LDP and CDP. In the  amplification \cite{erlingsson2019amplification} model they provide a differential privacy algorithm that will be satisfied by any $\epsilon$ - invariant algorithm in the LDP as well as in the CDP model.
\par Our approach is to keep the best characteristics from both the LDP and CDP. Here, 
\begin{itemize}
    \item We are taking the reports from the already industry standard RAPPOR algorithm.
    \item Then we calculate the tf-idf value for all the positions where at least one 'on bit' has occurred with respect to the sample size after sampling multiple times with different sample size.
    \item We get a constant \textit{attribution/ sample size} always after sampling for a constant number of occurrence of 'on bit' in the reporting string of prr and irr of RAPPOR.
    \item After that in the calculation step the weighted sum is being calculated. The results are then stored in a separate database. 
    \item This whole approach assures that no direct identification is possible in case of cohort, prr or irr by the curator of the database or any attacker.
\end{itemize}  
\par From our approach we contribute several things-
\begin{itemize}
    \item The storing database in the server side becomes less in size.
    \item We are not storing anything other than the weighted sum only.
    \item The analysis phase is faster.
    \item Our model identifies the major '\textit{True Value}' (which has the more occurrences in the samples) every time.
\end{itemize}
\par Our overall approach is to give a centralized environment to the RAPPOR LDP model that will eventually give a more generalized bigger picture to ensure more differential privacy in a comparative faster way which is better than the other centralized approach such as OUTIS\cite{chowdhury2019outis}, Amplification model\cite{erlingsson2019amplification} and PROCHLO\cite{bittau2017prochlo}. 

\par This paper is divided into six sections. After the introduction(in section-\ref{intro}) part, we will discuss about the background theories in the related work(in section-\ref{sec:2}) , then in the previous approach(in section-\ref{sec:3})  we will discuss about the the previous approaches that helps us to set our aim to achieve the current result. The last three sections all are devoted to talk about the architecture(in section-\ref{sec:4}), results and analysis (in section-\ref{sec:5}) about our approach that will conclude(in section-\ref{sec:6}) with our thought of future approach.
The description of the acronyms used in this paper is given in the following Table \ref{tab:ac}:
\begin{table}[ht]
\caption{ Name and description of all the acronyms used}
\label{tab:ac}       
\begin{tabular}{|p{6cm}|p{6cm}|}
\hline
\textbf{Name of the Acronyms}& \textbf{Description} \\
\hline
\textit{DP} & Differential Privacy. \\
\hline
 \textit{LDP} & Local Differential Privacy.\\
 \hline
 \textit{CDP} & Central Differential Privacy\\
 \hline
\textit{RAPPOR} & Randomized Aggregatable Privacy Preserving Ordinal Response.\\
 \hline
  \textit{ESA} &  Encode-Shuffle-Analyze\\
  \hline
   \textit{OUTIS} & A system for Crypto-Assisted Differential Privacy on Untrusted Servers.\\
   \hline
  

\end{tabular}
\end{table}

\section{Related Work}
\label{sec:2}
Differential privacy guarantees the following two things - \begin{itemize}
    \item  the output of the differential privacy algorithm is definitely stable and
    \item  only the forest of sampling and analysis is guaranteed which ensures that if the input of a single user get changed it will not have any effect on the output at all.
\end{itemize} 
\subsection{\textbf{Differential Privacy}} \cite{dwork2014algorithmic}“Differential privacy” describes a promise, made by a data holder or curator, to a data subject: “You will not be affected, adversely or otherwise, by allowing your data to be used in any study or analysis, no matter what other studies, data sets, or information sources, are available.” The mathematical definition after having the standard relaxation with an addition of $\delta$ will be as follows -
\par A randomized algorithm $A : D^n \to S$ satisfies \((\epsilon,\delta\)) - differential privacy if for all S $\subset$ S and for all adjacent $D$, $D' \in D$ it holds that 
\begin{equation}
    Pr[(A(D)\in S]\leq e^\epsilon Pr[A(D') \in S]+ \delta
\end{equation}
\par In Table \ref{tab:2} we have provided the essential explanation of the parameters in the equation (1).
\begin{table}[ht]
\caption{Description of all the parameters in the \textit{Differential Privacy}}
\label{tab:2}       
\begin{tabular}{|p{6cm}|p{6cm}|}
\hline
\textbf{Name of the Parameters}& \textbf{Description} \\
\hline
data holder or curator & The only trusted \textit{3rd party} person who has the whole
                       authority of the database. \\
\hline
 data subject & The producer of the data or the data owners.\\
 \hline
 $A$ & Randomized algorithm\\
 \hline
 $D$ and $D'$ & They are adjacent inputs and application-dependent. They must differ in one of the n elements.\\
 \hline
  $\delta$ &  It is a small constant and  is smaller than $1/|D|$.\\
  \hline
   $Pr$ & is the probability taken over after the coin tossing of $A$.\\
   \hline
   $S$ & the range of the randomized algorithm $A$\\
   \hline
   $\epsilon$ & Measure of how much privacy can be lost.\\
   \hline

\end{tabular}
\end{table}
 From the above definition and  \cite{kasiviswanathan2011can} it is evident that differential privacy is a good measure of the output stability in a randomized response technique when the input can change frequently from any single user.DP is of two types:
 
 \subsubsection{\textbf{Local Differential Privacy} (LDP)} This is entirely depended on \textit{randomized response} technique introduced in 1965 which is a simple response technique by the user depending on the coin toss probability. This model was formally introduced by kasiviswanathan et al.\cite{kasiviswanathan2011can} In this model the distribution of a forest of data is always assured to be stable even when a user can change his or her response suddenly. This simple trust model is the main attraction of this model which is why it is the main adoption in the industrial implementation now a days which does not even need a third party assurance at all. 
 
 \subsubsection{\textbf{Central Differential Privacy} (CDP)} Here the trusted data curator has the main role to play.The curator has the responsibility to add the uncertainty by adding random noise which eventually led to differential privacy. This whole process occurs entirely the untrusted data analyst's queries. As the answer to the queries always hold a small fraction of data of the whole centralized dataset which helps to establish the differential privacy phenomenon. 
 
 The comparison between \textit{CDP} and \textit{LDP} is in the Table \ref{tab:com}
 \begin{table}[ht]
    \caption{Comparison between LDP and CDP}\label{tab:com}
    \begin{tabular}{|p{6cm}|p{6cm}|}
    \hline
    LDP &  CDP \\
    \hline
    There is neither any trusted data curator nor any data preservation &  Individual clients provide the data and the data are stored in a trusted database.\\
    \hline
    It assures that the report is produced by individual client only where \textit{randomized response of coin} toss mechanism is used to produce the report. & The curator release the final report using the Laplace distribution of noise to hide the presence or absence of particular participant. \\
    \hline
    \textit{RAPPOR} is the current state of art mechanism for the google chrome search engine which is very fast. & In comparison to \textit{RAPPOR} it is slower and expensive to implement.\\
    \hline
    \end{tabular}
\end{table}
 
\subsection{ \textbf{Randomized Response Technique}} It is introduced in 1965 by Stanly L.Warner \cite{warner1965randomized}, it is a very necessary and effective technique in terms of structured survey, which helps the participants to answer freely about the sensitive issues without being concerned about confidentiality. The whole model is based on the participants’ truthfulness. It can be interpreted like following - 
\begin{equation}
\begin{split}
    YA & = p \times EP + (1-p)(1-EP) \\
    or,  EP & = (YA + p -1)/ (2p - 1)
\end{split}
\end{equation}

\begin{table}[ht]
\caption{Description of all the parameters in the \textit{Randomized Response Technique}}
\label{tab:3}       
\begin{tabular}{|p{6cm}|p{6cm}|}
\hline
\textbf{Name of the Parameters}& \textbf{Description} \\
\hline
$YA$ & The proportion of 'yes'- answer \\
\hline
$p$ & the probability to answer the sensitive question.\\
\hline
 $A$ & Randomized algorithm\\
 \hline
 $EP$ & the true proportion of those interviewed bearing the embarrassing property\\
 \hline
  

\end{tabular}
\end{table}
In Table \ref{tab:3} the description of the parameters are given. We can consider the following example- Let’s ask a number of cancer patient under clinical observation, whether they have consumed the medicine  today or not. They should answer ‘YES’ if it is a 6 and answer truthfully if it otherwise, after flipping a dice secretly which is unknown to the interviewer. Let the number of ‘YES’ is 55 out of 100 participants. Therefore $YA$  = 55/100= 11/20. Therefore p = 1/6. therefore $EP$  = 42.42\% or true proportion of medicine taker is 42.42\%.

\section{Previous Approaches}
\label{sec:3}
\subsection{\textbf{RAPPOR}} Randomized Aggregatable Privacy-Preserving Ordinal Response\cite{erlingsson2014rappor} or \textit{RAPPOR} is a industry standard open-source technology introduced by Google whose main function is to get the raw data \textit{locally} from the client software, anonymously with the guarantee of differential privacy. They have applied randomized response in a novel manner by the introduction of Bloom filter. \textit{RAPPOR} always make sure that the forest of client side strings can only get analyzed by the untrusted analyst but not the other way around. It is a very good example of LDP. After discussing CDP and it's implementation \textit{PROCHLO} in the next two paragraphs, the difference between CDP and LDP will be more clear. 

\subsection{\textbf{PROCHLO implementation of ESA}} This is a very good example of CDP. The entire architecture of ESA or \textit{Encode, Shuffle} and Analyze to be precised is a three-step pipeline. PROCHLO \cite{bittau2017prochlo} is the real life implementation of ESA, with the introduction of stash shuffle, which is a novel, scalable and efficient oblivious-shuffling algorithm using the Intel's SGX architecture. In this architecture there is a central database of encrypted records which can only be decoded by a special analysis algorithm determined by the corresponding decryption key. After that in the shuffling stage the data truly become the part of the crowd with the assurance of that a threshold number of reported items will surely be existed for every data partition. The analyze step aggregates all the data partition publicly though the anonymity always intact. In every step of ESA the differential privacy has been maintained \cite{bittau2017prochlo}. They have also taken the deep learning area into consideration and implement some test cases with success.
 Kashiviswanathan et al\cite{kasiviswanathan2011can} has stated that there exits an exponential separation between the sampling complexity and accuracy in the LDP and CDP algorithms. That's why LDP requires a huge amount of data to have a good population statistics. So the main gap between the CDP and the LDP approach is mainly of four kinds, like - \begin{itemize}
    \item differencein necessary amount of data to produce a good population distri-
bution statistics,
\item storage of data,
\item difference in speed,
\item last but not the least difference in approach.\end{itemize}
\par \textit{ESA} and its implementation \textit{PROCHLO}\cite{bittau2017prochlo} is a very good implementation of CDP. We have already discussed it above.
\par As we have discussed before there have been some model proposed in recent literature, like - OUTIS\cite{chowdhury2019outis}, Amplification by shuffling \cite{erlingsson2019amplification} etc. They have their own pros and cons.   
 \subsection{\textbf{Amplification Approach}} In \cite{erlingsson2019amplification} erlingsson et al. have proposed an algorithm that gives a powerful amplificaiton technique, that any permutation invariant algorithm satisfying $\epsilon$will satisfy $\mathcal{O}(\epsilon\sqrt{\log{(1/\delta)/n}},\delta)$ - central differential privacy. But their assumption standardized on the basis of static population, which is not possible in real life. Also they have ignored the implications of timing or traffic channels, which should have been considered.This is a huge drawback comparison to the \textit{ESA} architecture. 
 \subsection{\textbf{OUTIS}} \textit{OUTIS} replaces a single trusted data curator by two untrusted non-collaborative servers i.e. the CSP and the AS the third party association in the CDP model is diminished. It gives the permission to the analysts to author the logical programs as logical programs always support the differential privacy, by restricting the access of the sensitive data.  
 \par The main cons of this model are -
 \begin{itemize}
     \item Aggregation operators, Multi-table queries, Matrix multiplication does not work in this environment.
     \item These servers are semi-honest ( this is achieved by linear homomorphic encryption and Yao's garbled circuits) which means they follow the protocols honestly but their contents and computation can be observed by an adversary.
     \item Privacy engine is not a strong point of this model.
     \item Too much work pressure on the AS part
     \item It starts off with a total privacy budget of $\epsilon^B$ agreed upon by all the data owners. So, there is an option that the privacy is not enough.
 \end{itemize}
 \subsection{\textbf{Initial Approaches}} Our \textit{first} approach was to make an environment where both the \textit{RAPPOR} and \textit{ESA} will be implemented and will be chosen as per as the query requirement. But for that approach, we have to implement the expensive \textit{ESA} architecture parallel to the \textit{RAPPOR} algorithm. Also the query should be known beforehand. That is why we did not implement this idea physically.
\par Our \textit{second} approach was to use a \textit{Convolutional Neural Network} layer on top of the \textit{RAPPOR} reports, which will eventually analyse the reports taking combination of prr and irr \cite{erlingsson2014rappor} as training data in sets of samples. After training, the model was not able to detect the true values beyond 2\% of accuracy which was really poor. The main reason behind our failure was mainly for overfitting due to the inbuilt noise from randomized response sampling in the prr and irr combination.
\par Our target from the beginning is to build an analysis technique which will follow the CDP, as well as easy to understand, robust, fast and cheap in implement. All these points have also been discussed in the previous sections too. That's why our \textit{third} approach is to follow the already renowned \textit{TF-IDF} technique (mainly used in the information retrieval field for relevant decision making)\cite{ACMB033AHCWU}. It is very easy to implement and as it follows the probabilistic relevance of 'term's (in our case 'on bit' and 'off bit' in the prr and irr string) in a document as well as in a corpus of documents the analysis part became very easy from our side. The details of our proposed architecture is being discussed in the following section \ref{sec:4}. 
 \section{The ARA Architecture}
 \label{sec:4}
 Our aim is to build a CDP system that is cheap, fast, robust and less complex. So, after two failed attempts we were success full to achieve at least a part of what have we aimed. The following four subsections will explain the methodology of our architecture, the data collection part and how both these parts has built the system.
 \subsection{\textbf{Tf-Idf}} The full form of \textit{Tf-Idf} is term frequency-inverse document frequency \cite{aizawa2003tf-idf}. It is mostly used to retrieve the " probability-weighted amount of information " mainly in the field of feature extraction of machine learning, automatic term extraction in computational terminology, information theoretic field, relative decision making etc. Term Frequency measures the "local relevance" \cite{ACMB033AHCWU} in a specific document of a corpus of documents. So, it provides a direct estimation of the probability of occurrence of a specific text or word after normalization with respect to the scope of calculation. Inverse Document Frequency on the other hand provide wide relevance in the whole document. The formula to calculate TF is following-
\begin{equation}
   TF(t,d) = f\textsubscript{t,d} / (number\:of\: words\: in \: d)
\end{equation}
where, $t$ is the term and $d$ is the document.
\par The formula to calculate IDF is following -
\begin{equation}
    IDF(t,d) = \log{\frac{N}{1 + \left|{d\epsilon D: t \epsilon d}\right| }}
\end{equation}
Where, $N$ = Total number of documents in the corpus and $\left|{d\epsilon D: t \epsilon d}\right|$ is the number of documents where $t$ appears. If the term is not in the corpus, then the IDF term will be undefined, that's why it is being adjusted by adding '1' in the denominator.
\par Therefore the total calculation is - 
\begin{equation}
    \textit{TF-IDF} = TF(t,d) \times IDF(t,d)
\end{equation}
The explanation of all the parameters that have been used in the calculation of \textit{TF-IDF} is being listed in the Table \ref{tab:4}.
\begin{table}[ht]
\caption{Description of all the parameters in the \textit{TF-IDF} calculation}
\label{tab:4}       
\begin{tabular}{|p{6cm}|p{6cm}|}
\hline
\textbf{Name of the Parameters}& \textbf{Description} \\
\hline
$TF(t,d)$ & Term Frequency \\
\hline
$t$ & the required term\\
\hline
 $d$ & Document in which the term $t$ has occurred\\
 \hline
 $f\textsubscript{t,d}$ & frequency of occurrence $t$ in $d$\\
 \hline
 $IDF(t,d)$ & Inverse document frequency\\
 \hline
 $N$ & Total number of documents in the corpus\\
 \hline
 1 + $\left|{d\epsilon D: t \epsilon d}\right|$ & Number of documents where $t$ has appeared\\
 \hline
  

\end{tabular}
\end{table}
 \subsection{Methodology}In our proposed \textit{ARA} analysis model the actions are divided into three steps. In the first step or in the \textit{Sampling} step we have sampled the reports taken from \textit{RAPPOR} and calculated the \textit{TF-IDF} value with the following formula for the 'prr' and 'irr' string -
 \begin{equation}
     \textit{TF-IDF} = \frac{N}{S} \times \log_{10} \frac{32}{S}
 \end{equation}
 Here $N$ = count of 'on bit' in the string and $S$ = Sample Size.
 \par After sampling (taking 100, 1000, 10000, 20000, 25000 samples at a time) from the reports many times we could deduce the following two important decisions-
 \begin{itemize}
     \item There are always a constant contribution from the 'on bit's to the \textit{TF-IDF} value abiding the $\frac{constant\:value}{Sample\:Size}$ rule depending on the number of 'on bit' in the string, for the whole sample size..
     \item As the size of our string is 32 bits, we also have deduced that there could not be more than 17 'on bit' in a string.
 \end{itemize}
\par The list of the constant with respect to the number of 'on bit' in a string is in Table \ref{tab:5}
\begin{center} 
\begin{table}[ht]
\caption{List of Constant contribution with respect to the number of 'on bit' in a string}
\label{tab:5}       

  \begin{tabular}{|p{6cm}|p{6cm}|}
\hline
\textbf{Number of 'on bit' in the string}& \textbf{Constant Value $C\textsubscript{v}$ ($v$ ranged from 1 - 17)} \\
\hline
1 & 1.20201279 \\
\hline
2 & 1.0927389\\
\hline
 3 & 0.993399\\
 \hline
 4 & 0.90309\\
 \hline
 5 & 0.80618\\
 \hline
 6 & 0.727\\
 \hline
 7 & 0.660052\\
 \hline
 8 & 0.60206 \\
 \hline
 9 & 0.550907 \\
 \hline
 10 & 0.50515 \\
 \hline
 11 & 0.4637573 \\
 \hline
 12 & 0.425969\\
 \hline
 13 & 0.3912066\\
 \hline
 14 & 0.3590219 \\
 \hline
 15 & 0.329059\\
 \hline
 16 & 0.30103\\
 \hline 
 17 & 0.274701\\
 \hline
  

\end{tabular}
\end{table}
  
\end{center}
These constants values are being kept secret and the database curator or the analyst only has the access to these constants. Also these constants are 1.1 times larger whenever the number of position gets lesser by one position. 
\par In the second step or in the \textit{Weighted Sum} calculation step we simply calculate the weighted sum for each report consisting of cohort value, prr, irr and true value using the following formula (Description of parameters used in weighted sum calculation step has been given in the Table \ref{tab:7}) -
\begin{equation}
    W = [ C\textsubscript{prr}\times C\textsubscript{C\textsubscript{prr}} + C\textsubscript{irr} \times C\textsubscript{ C\textsubscript{irr}} ] \times V
\end{equation}
\par Here, $W$ = weighted sum, $C\textsubscript{prr}$ = the count of 'on bit' in the prr string, $C\textsubscript{irr}$ = the count of 'on bit' in the irr string, $V$ = cohort value, $C\textsubscript{C\textsubscript{prr}}$ = constant value from the table \ref{tab:5},$C\textsubscript{C\textsubscript{irr}}$ = constant value from the table \ref{tab:5} . The description of \textit{prr, irr, cohort} is given in the \cite{erlingsson2014rappor}. If, $V$ = 0 then the formula will be following -
\begin{equation}
    W = [ C\textsubscript{prr}\times C\textsubscript{C\textsubscript{prr}} + C\textsubscript{irr} \times C\textsubscript{ C\textsubscript{irr}} ]
\end{equation}

\begin{table}[ht]
\caption{Description of parameters used in weighted sum calculation step}
\label{tab:7}       
\begin{tabular}{|p{6cm}|p{6cm}|}
\hline
\textbf{Name of the Parameters}& \textbf{Description} \\
\hline
$W$ & Weighted Sum \\
\hline
$C\textsubscript{prr}$ & the count of 'on bit' in the prr string\\
\hline
 $C\textsubscript{irr}$ & the count of 'on bit' in the irr string\\
 \hline
 $V$ & cohort value\\
 \hline
 $C\textsubscript{C\textsubscript{prr}}$ & constant value from the table \ref{tab:5}\\
 \hline
 $C\textsubscript{C\textsubscript{irr}}$ & constant value from the table \ref{tab:5}\\
 \hline
  

\end{tabular}
\end{table}

\par  . For experimental convenience, we have used 10 true values starting from v1, v2, v3, ..., v10 and the range of cohort is 0 - 63. The calculated weighted sum are then stored in a centralized database. Point to be noted that there are only two attributes that we are storing - the \textit{constant values} and the \textit{weighted sums}. Therefore there is no direct identification of the reports that is being stored. So, the security is not harmed, as there are so many combination of count of 'on bit' in prr and irr with respect to the true values and cohort.
\par
\begin{figure*}[ht]
\includegraphics[width=0.80\textwidth]{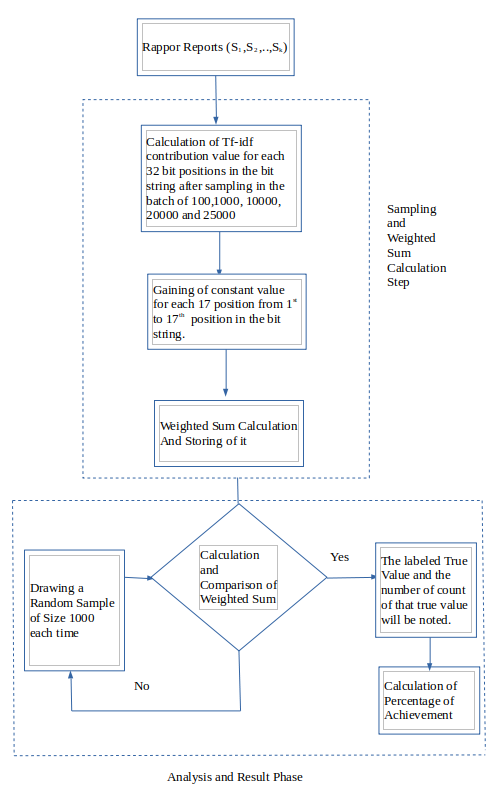}
\centering
\caption{Data flow in the ARA architecture}
\label{fig:df}       
\end{figure*}

 The third step is the last step. It is addressed as the \textit{Analysis Phase}. Here we take the testing samples of reports and then calculate the weighted sum using the same formulas above and match them against the central database and generate a report. The Data Flow in the proposed system is showed in the Fig. \ref{fig:df}

\subsection{Data}
\begin{figure*}[ht]
\includegraphics[width=0.80\textwidth]{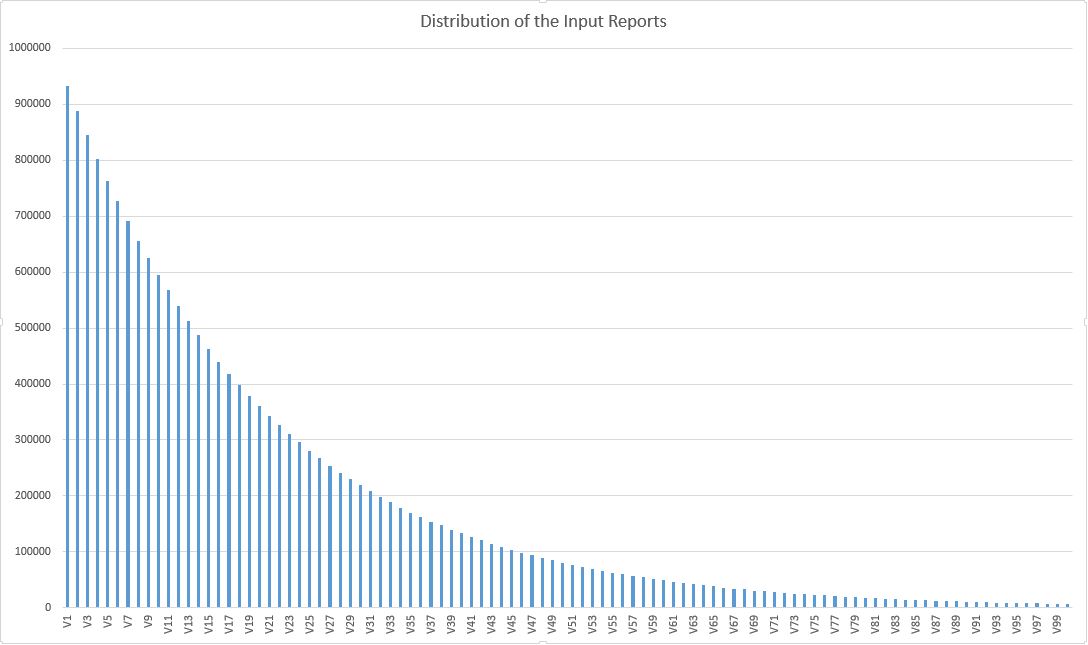}
\centering
\caption{Distribution of the Input reports arranged from \textit{RAPPOR}}
\label{fig:1}       
\end{figure*}

 We collected data after cloning the \textit{RAPPOR} implementation in the Google repository from GitHub \cite{GitHub2017Rappor} and running it more than hundred times. We made our own datasets from the generated reports from running. For our convenience we used only ten true values (v1, v2, v3, ..., v10) where \textit{RAPPOR} has used one hundred true values (v1, v2, v3, ..., v100). The distribution of the input report is being shown in the Figure \ref{fig:1}.
As we can see the reports are already distributed abiding the exponential distribution. So, there is already sufficient noise implemented inside the reports. As it is an open source repository we do not need to fetch the permission of authors to use these codes. 
\subsection{System Specification}
Experiments were run on a Desktop Computer with an Intel(R) Core(TM) i7-7700 CPU  running at 3.60GHz using 8192 KB of cache memory on the Linux 18.04 UBUNTU OS with 1000 GB of HDD. It is a 64 bit system. It took around 1.5 hour to complete one test set at a time. We wrote all our code using R Studio, where R version 3.5.1 (2018-07-02) -- "Feather Spray" on the Platform: x86\_64-conda\_cos6-linux-gnu (64-bit) is used. All our codes have been uploaded here \cite{GitHub2019ARA}.
\begin{center}
\begin{table}[ht]
\caption{Summary of the Test Results}
\label{tab:6}       
\begin{tabular}{ |p{3cm}|p{3cm}|p{3cm}|p{3cm}|}
\hline
 \multicolumn{4}{|c|}{Total Sample Size 1000 }  \\
\hline
Test No. & Major True Value/ Component & Sample Limit & Percentage of Achievement  \\
            
            \hline
           1 &  v1  & 30 - 300  &        46      \\
           \hline
           2            &      v1 &  10 - 400        &  93.35    \\
           \hline
           3           &       v1    &  0 - 500     &   72.2    \\
           \hline
           4           &       v1   &  0 - 600      &  51.8     \\
           \hline
            5           &        v2 &  0 - 300       &  32.4     \\
            \hline
           6 &  v2  & 5 - 400         & 59.4      \\
           \hline
           7           &       v2  &   0 - 500      &  31.33     \\
            \hline
           8           &       v2   &  0 - 600     & 67.64      \\
           \hline
           9           &       v3  &  0 - 300       & 60.9      \\
            \hline
           10          &       v3 &   0 - 400        &   40.1    \\
            \hline
           11           &      v3  &  5 - 500        &   43.7    \\
            
            \hline
           12           &       v3    &  0 - 600     &  59.1     \\
           \hline
           13            &       v4 &  50 - 300       & 33.1      \\
            \hline
           14           &      v4  &  5 - 400        & 42.74      \\
           \hline
           15            &       v4  &  0 - 500      &  87.9     \\
            
           \hline
           16          &        v4 &  50 - 500       &  60.12     \\
           
            \hline
           17          &       v4   &  0 - 600       &  68     \\
           
           \hline
           18           &       v5 &  50 - 300         &  43.1     \\
            \hline
           19           &        v5   &  0 - 400     & 49.82      \\
            
            \hline
           20            &       v5  &  50 - 500     &  40.11     \\
           \hline
           21          &     v5  &   5 - 600         &   53    \\
            \hline
           22            &       v6  &  50 - 300     & 40.17      \\
            \hline
           23           &       v6  &  50 - 400      & 63.4       \\
           \hline
           24          &      v6  &   0 - 500        &   41.7    \\
            \hline
           25           &       v6  &  20 - 600      & 53.1      \\
           \hline
           26           &      v7  &   5 - 300       & 46.8      \\
            \hline
           27           &       v7  &  0 - 400       &  45.2     \\
            
            \hline
           28           &       v7  &  0 - 500       &  46.5     \\
            \hline
           29          &        v7  &  0 - 600      &  65.53     \\
            \hline
           30            &       v8   &  40 -300     & 46.3      \\
            \hline
           31            &      v8   &  0 - 400      & 32.8      \\
            \hline
           32           &       v8   &  0 - 500      & 63.5      \\
            \hline
           33           &       v8 &  0 - 800        &  55.9     \\
            \hline
           34           &      v9  &  10 - 400        &  60.3     \\
            \hline
           35          &       v9  &  0 - 500        & 60.96      \\
            \hline
           36           &       v9   &  5 - 600      &  66.92     \\
           
           \hline
           37           &      v10 &   20 - 300        &  36.8     \\
            \hline
           38         &     v10    &    20 - 400       & 37.8      \\
            
             \hline
           39         &     v10 &   5 - 500           & 51.52      \\
            \hline
           40            &      v10   &  0 - 600      & 40.1      \\
            \hline

\end{tabular}
\end{table}
\end{center}
   \section{Result and analysis}
\label{sec:5}
 
\subsection{Result}
For testing our model we have taken 1000 sample reports at a time consisting of the cohort, prr and irr strings. We then test the set of reports for 100 times each, against our model. We have done total 40 set of tests. After gaining the count of the occurrence of a true value in the matching process which is being described in the flow chart, the percentage achievement is simply the percentage calculation of the count against the sample size. The summary of the test results is in the Table \ref{tab:6} .

\subsection{Analysis} 
\begin{figure*}[ht]
\includegraphics[width=0.80\textwidth]{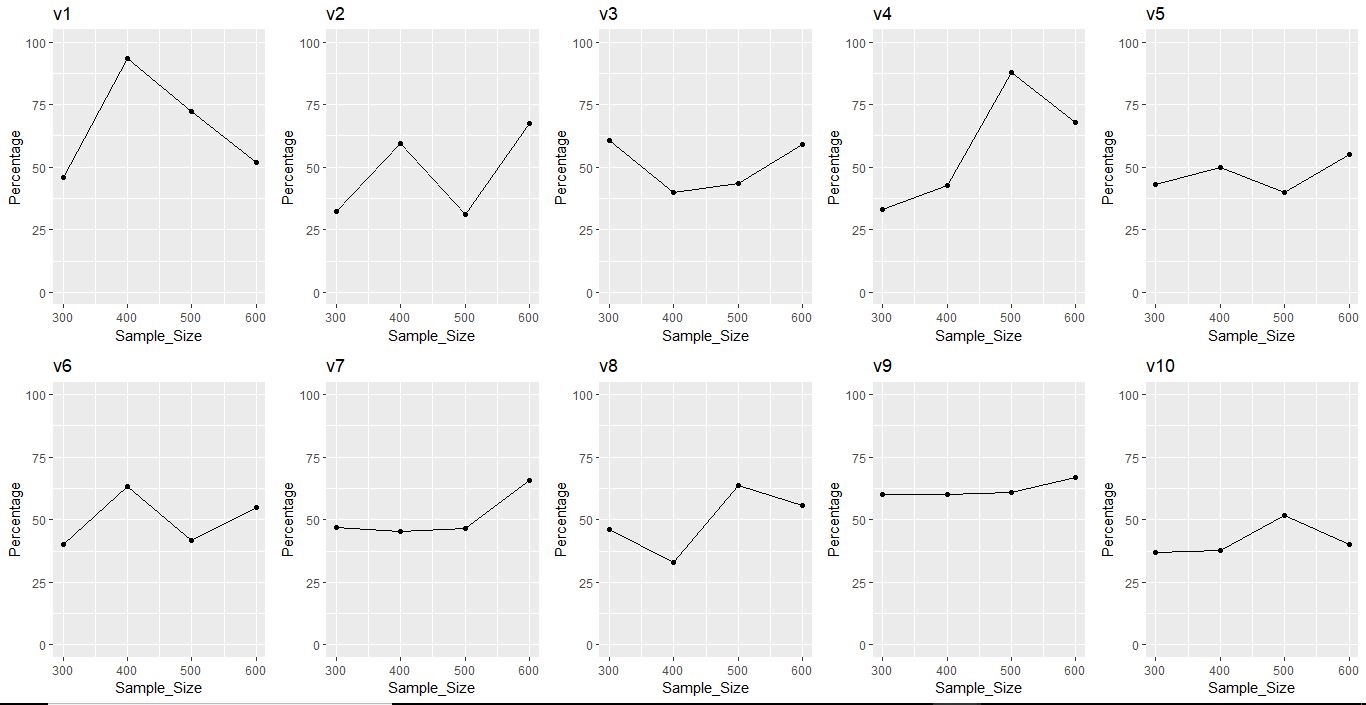}
\centering
\caption{An Analysis between sample size and percentage of achievement}
\label{fig:2}       
\end{figure*}

From the test part it is evident that, every time our model was able to detect the major component out of the samples which has been collected from multiple clients at a time. Though the percentage of achievement is an issue, but still our model is fast. Also, the percentage of achievement and the sample size is not depended on each other. It is being shown in the following graphical representation in Figure \ref{fig:2} -

\subsection{Comparison}
We have outlined a comparison of our approach against current existing central approaches \textit{ESA}, \textit{OUTIS} and \textit{Amplification Model} in Table \ref{tab:comp}. However, a direct comparison is not possible; given some of the approaches are hardware approaches and some are purely theoretical whose implementations are rather expensive and unavailable or currently do not exist respectively. \\

\begin{table}
\caption{Comparison with other models}
\label{tab:comp}       
\begin{tabular}{|p{6cm}|p{6cm}|}
\hline
\textbf{Name of the Models}& \textbf{Execution Time} \\
\hline
ESA & 4.1 hour for Encoder + Shuffler 1 for 10 million input \\
\hline
OUTIS & Approximately 8 hours for Adult dataset from University of California, Irvine repository, with 32,651 data in case of unoptimized OUTIS.\\
\hline
 Amplification model & Currently a theoretical approach. No current implementation is available.\\
 \hline
 ARA & Approximately 1.25 hours for one set of complete analysis phase which consists of 1000 randomly selected samples and 100 times matching test against the samples.\\
 \hline


\end{tabular}
\end{table}
\section{Conclusion}
\label{sec:6}
Privacy has been a long-established issue through decade. Though Differential Privacy has paved a significance contribution in this area but the achievement in the Local differential privacy area is still better in context of industry standard, speed, utility, expensiveness etc. Our model is highlighted on these issues.The goodness of our model are-
\begin{itemize}
    \item It is fast in the analysis phase.
    \item The Centralized Database size is much smaller.
    \item The database does not store the reports for longer. Just the time for the calculation of weight only.
    \item Accurately identify the major true value every time.
    \item Simple probabilistic approach towards analysis, which is not as complex as \textit{OUTIS} or \textit{PROCHLO}.
    \item It maintains \textit{RAPPOR}'s differential privacy promises.
\end{itemize}
The main drawback we get is the level of achievement is not more than 52.28\% in average. Also our model is not able to detect the second major component with accuracy too. The utility and flexibility should be more too.

The drawbacks are definitely the motivation for our future work. But it is a simple approach towards centralized differential privacy which is less complex in framework and have an accessible computation. Therefore it is in turn definitely a good contribution towards central differential privacy.  

\section{Conflict of Interest} :
 The authors declare that they have no conflict of interest.

\end{document}